\documentclass{ws-procs961x669}            
\begin{document}
\title{Alternatives to $\Lambda$: Torsion, Generalized Couplings, and Scale Invariance}

\author{C. J. A. P. Martins$^*$}
\address{Centro de Astrof\'{\i}sica da Universidade do Porto, and\\
Instituto de Astrof\'{\i}sica e Ci\^encias do Espa\c co, Universidade do Porto,\\
Rua das Estrelas, 4150-762 Porto, Portugal\\
$^*$E-mail: Carlos.Martins@astro.up.pt}

\author{C. M. J. Marques}
\address{Faculdade de Ci\^encias e Tecnologia, Universidade Nova de Lisboa,\\ 2829-516 Caparica, Portugal, and\\
Centro de Astrof\'{\i}sica da Universidade do Porto,\\
Rua das Estrelas, 4150-762 Porto, Portugal}

\author{C. B. D. Fernandes, J. S. J. S. Oliveira, D. A. R. Pinheiro and B. A. R. Rocha}
\address{Faculdade de Ci\^encias, Universidade do Porto,\\
Rua do Campo Alegre, 4150-007 Porto, Portugal, and\\
Centro de Astrof\'{\i}sica da Universidade do Porto,\\
Rua das Estrelas, 4150-762 Porto, Portugal}

\begin{abstract}
We present a comparative analysis of current observational constraints on three recently discussed alternative models for explaining the low-redshift acceleration of the universe: the so-called steady-state torsion model, the generalized coupling model, and the scale invariant model by Maeder (an example of a broader class which we also briefly study) These are compared to the traditional parameterization of Chevallier, Polarski and Linder. Each of the candidate models is studied under two different assumptions: as genuine alternatives to $\Lambda$CDM (where a new degree of freedom would be expected to explain the recent acceleration of the universe without any cosmological constant) and as parametric extensions of $\Lambda$CDM (where both a cosmological constant and the new mechanism can coexist, and the relative contributions of both are determined by the data). Our comparative analysis suggests that, from a phenomenological point of view, all such models neatly divide into two classes, with different observational consequences.
\end{abstract}


\bodymatter

\section{Introduction}

The observational evidence for the acceleration of the universe shows that our canonical theories of cosmology and particle physics are at least incomplete, and possibly incorrect. Is dark energy a cosmological constant (i.e. vacuum energy)? If the answer is yes, it is ten to some large power times smaller than our Quantum Field Theory based expectations. If the answer is no, then the Einstein Equivalence Principle must be violated. Either way, new physics is out there, waiting to be discovered; we must search for, identify and characterize this new physics. The CosmoESPRESSO team uses the universe as a laboratory to characterize, with precision spectroscopy and other observational, computational and theoretical tools, the behaviour of the gravitational interaction, with he goal of determining what makes the universe accelerate. In what follows we highlight recent contributions of the CosmoESPRESSO team to this fundamental quest.

The search for the physical mechanism underlying the observed low-redshift acceleration of the universe is the most compelling goal of modern fundamental cosmology, and several theoretical possibilities beyond a cosmological constant can be envisaged in principle, each with its specific observational consequences.

Our goal here is to present a comparative study of the observational constraints on three classes of alternative models: the so-called steady-state torsion model \cite{Kranas} of Kranas \textit{et al.}, the generalized coupling model \cite{Feng} of Feng and Carloni, and the scale invariant model \cite{Maeder} of Maeder; the latter is an example of a broader class of models \cite{Canuto1,Canuto2} first proposed by Canuto \textit{et al.}, which we also briefly study. As a benchmark we use the traditional phenomenological parameterization of Chevallier, Polarski and Linder (henceforth CPL) \cite{Chevallier,Linder}. All models have common parameters (specifically, the matter density parameter, $\Omega_m$) but also some specific ones, and a comparative analysis using a common data set is therefore interesting.

We take three models at face value and phenomenologically constrain them through a standard likelihood analysis using low-redshift background cosmology data. Specifically, we use the recent Pantheon dataset \cite{Riess}, including its covariance matrix. We also use a compilation of 38 Hubble parameter measurements \cite{Farooq}. Occasionally we will also use a Planck prior \cite{Planck} on the matter density, $\Omega_m=0.0315\pm0.007$. The value of the he Hubble constant is always marginalized analytically, following the procedure detailed in Ref. \citenum{Basilakos}. The analysis is done on a grid (since we are only dealing with background cosmology, there is no computational need for a full MCMC analysis), and we have explicitly verified that the grid sizes that have been used are sufficiently large for the results presented in the following sections not to be affected by these sizes. Moreover, the following section will also present an explicit validation test of our code for the supernova data.We will work in units where the speed of light is set to $c=1$.

\section{Preamble: The CPL parameterization}

In the CPL parameterization the dark energy equation of state parameter is assumed to have the form \cite{Chevallier,Linder}
\begin{equation}
w(z)=\frac{p(z)}{\rho(z)}= w_0+w_a\frac{z}{1+z}\,,    
\end{equation}
where $w_0$ is its present value while $w_a$ quantifies its possible evolution. This is manifestly phenomenological: it is not intended to mimic a particular dark energy model, but aims to describe generic departures from the $\Lambda$CDM behaviour (which corresponds to $w_0=-1$ and $w_a=0$). In principle it allows for both canonical and phantom fields, since there is no restriction on the two model parameters, at least on purely mathematical grounds.

We assume a flat Friedmann-Lema\^{\i}tre-Robertson-Walker model, in which case the Friedmann equation has the form
\begin{equation}
\frac{H^2(z)}{H_0^2}=\Omega_m(1+z)^3+(1-\Omega_m)(1+z)^{3(1+w_0+w_a)}\exp{\left(-\frac{3w_az}{1+z}\right)}\,,
\end{equation}
where the matter parameter is $\Omega_m \equiv \kappa \rho_0 / 3 H_0^2$ and $\kappa=8\pi G$. This can now be constrained using the aforementioned data.

The case of a constant equation of state parameter (i.e. $w_a=0$), for the case of the supernova dataset, have been used in Ref. \citenum{Fernandes} as a validation test of our analysis code, against the results of Ref. \citenum{Riess}. In this case the one-sigma constraints on the two model parameters from the combined data sets are
\begin{eqnarray}
\Omega_m&=&0.27\pm0.02\\
w_0&=&-0.92\pm0.06\,,
\end{eqnarray}
which are compatible with $\Lambda$CDM.

For the full three-parameter CPL model, the one-sigma constraints on the three model parameters from the combined data sets are
\begin{eqnarray}
\Omega_m&=&0.26^{+0.03}_{-0.05}\\
w_0&=&-0.92^{+0.09}_{-0.08}\\
w_a&=&0.86^{+0.14}_{-0.24};
\end{eqnarray}
the reduced chi-square at the best fit is $\chi^2_\nu\sim0.9$, so the model is slightly overfitting the data (a behaviour which is mainly driven by the Hubble parameter data). The first two of these constraints are compatible with the values for the $w_0$CDM analysis (with naturally larger uncertainties), but there is a clear preference for a positive slope $w_a>0$. However there are strong degeneracies between the parameters, and the constraints do depend on the choice of priors. In the above we used the uniform prior on the matter density $\Omega_m=[0.05,0.5]$, the choice being motivated by the aforementioned validation of our code. As an illustration of the sensitivity of our results to this choice, if instead one uses the narrower uniform prior $\Omega_m=[0.15,0.45]$, one find
\begin{eqnarray}
\Omega_m&=&0.26^{+0.03}_{-0.05}\\
w_0&=&-0.92^{+0.07}_{-0.08}\\
w_a&=&0.74^{+0.21}_{-0.48};
\end{eqnarray}
in other words, there is no impact on the matter density and $w_0$, but there is a significant impact on $w_a$. Breaking these degeneracies requires additional data, for example from cosmic microwave background observations. In any case, our purpose here is to set up a benchmark for the constraining power of these data sets, against which to compare the constraints on the alternative models to be discussed in what follows.

\section{Steady-state torsion}

A possible extension of General Relativity consists in allowing for the presence of spacetime torsion. In such theories there is a further degree of freedom (in addition to the usual metric), which also gravitates. Mathematically, the torsion tensor is defined as the antisymmetric part of the affine connection; the symmetric part of the connection are the usual Christoffel symbols. Physically, this defines relation between the intrinsic angular momentum (i.e., the spin) of matter with the geometric properties of the underlying spacetime. The only non-trivial contraction of the torsion tensor is a torsion vector, and the general field equations including torsion are known as the Einstein-Cartan equations. Nominally the Einstein equations retain the usual form, but the presence of torsion implies that the Ricci tensor and the energy-momentum tensor are not symmetric. The Cartan equations relate the torsion tensor to the spin tensor, and similarly for the torsion and spin vectors.

The form of the underlying torsion tensor can be chosen such that the homogeneity and isotropy of FLRW universes is preserved \cite{Tsamparlis}, and in this case the remaining degree of freedom is a scalar function $\phi$ which must depend only on time (a spatial dependence would violate the homogeneity assumption), but is otherwise arbitrary. Making the standard assumption of treating the metric and the torsion as independent objects and furhter assuming a flat universe, one finds the following Friedmann, Raychaudhuri and continuity equations \cite{Kranas}
\begin{eqnarray}
H^2&=&\frac{1}{3}\kappa\rho\frac{1}{3}\Lambda-4\phi^2-4H\phi\\
\frac{\ddot a}{a}&=&-\frac{\kappa}{6}(\rho+3p)+\frac{1}{3}\Lambda-2{\dot\phi}-2H\phi\\
{\dot\rho}&=&-3H\left(1+2\frac{\phi}{H}\right)(\rho+p)+4\phi\left(\rho+\frac{\Lambda}{\kappa}\right)\,.
\end{eqnarray}
Here the dot denotes a derivative with respect to physical time, $H={\dot a}/a$ is the Hubble parameter, and $\rho$ and $p$ are the density and pressure. In what follows we will assume barotropic fluids with a constant equation of state $p=w\rho$. It has been recently suggested that such universes may undergo accelerating phases \cite{Kranas}. We can conveniently define a torsion contribution
\begin{equation}
\Omega_\phi=-4\left(\frac{\phi_0}{H_0}\right)\left[1+\frac{\phi_0}{H_0}\right]\,.
\end{equation}
In Ref. \citenum{Marques} these models were constrained under the so-called steady-state torsion assumption of a constant fractional contribution of torsion to the volume expansion, that is $\phi/H=\lambda=const.$.

It is easy to find, in agreement with other recent works, that models without a cosmological constant (where torsion itself would be expected to yield the current acceleration of the universe) are strongly disfavoured by the data. Indeed, in this case, for which the matter density would be given by $\Omega_m=(1+2\lambda)^2$, the best fit parameters would have a reduced chi-square of at least 2.7 for the datasets under consideration.

\begin{figure}
\begin{center}
\includegraphics[width=\columnwidth,keepaspectratio]{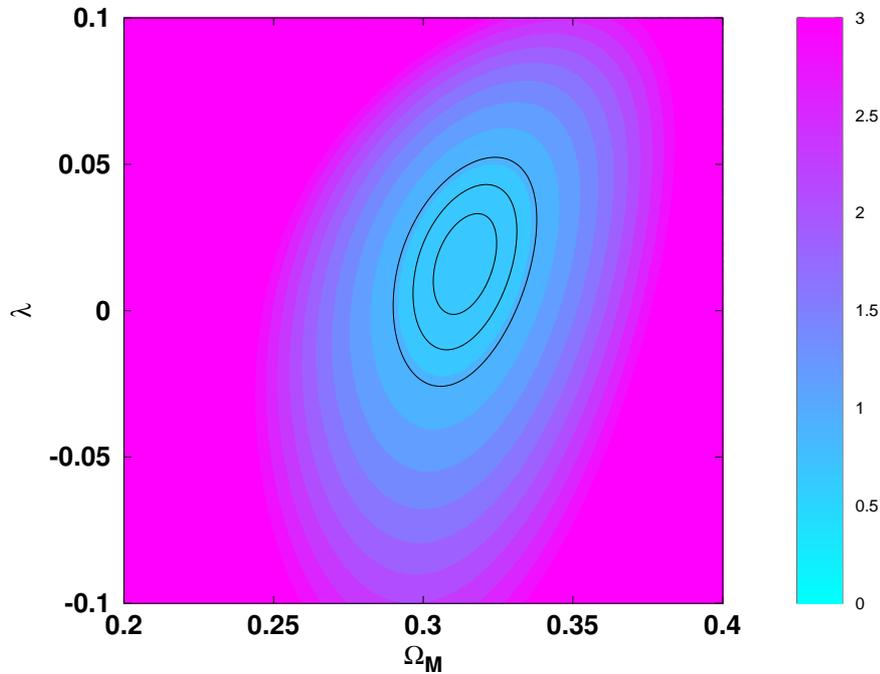}
\end{center}
\caption{Constraints on the $\lambda$--$\Omega_m$ parameter space for $w=0$. The black lines represent the one, two and three sigma confidence levels, and the colormap depicts the reduced chi-square of the fit, with points with $\chi^2_\nu>3$ shown in purple. Similar constraints can be found in Figure 2 of Ref. \citenum{Marques}.}
\label{fig1}
\end{figure}

However, one can also treat these models as one-parameter extensions of $\Lambda$CDM, whereby one can constrain the relative contributions of the cosmological constant and of torsion. In this case, if one assumes that that matter has the standard equation of state, $w=0$. As in the previous subsection will separately consider the cases without and with the aforementioned Planck prior on the matter density. Without the Planck prior, we find the following one-sigma posterior likelihoods for the two free parameters
\begin{eqnarray}
\lambda_{w=0}&=&-0.07^{+0.05}_{-0.04}\,\\
\Omega_{m,w=0}&=&0.18^{+0.06}_{-0.03}\,;
\end{eqnarray}
there is a clear degeneracy between the two parameters, and the preferred value of the matter density is lower. The inclusion of the Planck prior breaks the degeneracy and significantly improves the constraints, as shown in Fig. \ref{fig1}; the one-sigma posterior likelihood for the torsion parameter becomes
\begin{equation}
\lambda_{(w=0, Planck)}=0.02^{+0.01}_{-0.02}\,,
\end{equation}
which is consistent with the null result at just over one sigma. 

Allowing for a non-zero (but still constant) equation of state, there is a weak degeneracy between $w$ and the other model parameters, so although the constraints become weaker (as they must), both parameters are still well constrained by the data, provided the Planck prior is included. In this case we have
\begin{eqnarray}
\lambda_{(w\neq0, Planck)}&=&-0.01\pm0.02\,\\
w_{Planck}&=&-0.05\pm0.03\,;
\end{eqnarray}
compared to the $w=0$ case the best-fit value has changed sign, and the constraint is now consistent with the null result at one sigma.

Overall, we find no statistically significant preference for the presence of torsion. By itself it can't be responsible for the acceleration of the universe, and even if taken as an extension of the canonical $\Lambda$CDM paradigm the overall contribution to the Universe's energy budget is constrained to be no larger than a few percent. We also note that our constraints should be seen as conservative: an analysis including a full treatment of the cosmic microwave background should lead to stronger constraints.


\section{Generalized couplings}
\label{sect3}

The precise nature of the coupling between matter and the metric in the Einstein equations is at questionable assumption of the theory. One may therefore explore the possibility that this coupling is nontrivial. On such example is the Feng and Carloni's generalized coupling model \cite{Feng}, which is equivalent to General Relativity in vacuum, but still allows for a different behaviour within a matter distribution.

In this case the Friedmann and Raychaudhuri equations, assuming a flat universe, can be written \cite{Feng}
\begin{eqnarray}
3qH^2&=&\frac{256\kappa(1-pq)^3(q\rho + 1)^2}{[4+q(\rho-3p)]^4} + q\Lambda - \kappa\\
6q(\dot{H} + H^2) &=& \frac{256\kappa(1-pq)^3(q\rho + 1)[2-q(\rho+3p)]}{[4+q(\rho-3p)]^4} + 2(q\Lambda - \kappa)\,,
\end{eqnarray}
where $q$ is a model-specific parameter defined as $q=\kappa/\lambda$ (where $\lambda$, not to be confused with the analogous torsion parameter, is interpreted as being akin to the vacuum energy density generated by matter fields) and $p$ is the pressure of a fluid that is assumed to be barotropic, with an equation of state $p=w\rho$, where $w$ is a constant equation of state parameter. The corresponding continuity equation takes the form
\begin{equation}
    \dot{\rho} = -\frac{3H\rho(w+1)[q^2\rho^2w(3w-1)+q\rho (1-7w) + 4]}{q^2 \rho^2w(3w-1) - q\rho(3w^2+13w+2)+4}\,.
\end{equation}
Note the model is effectively a bimetric theory \cite{Feng}. In what follows we take the model as a phenomenological one and treat $q$ (or a dimensionless version thereof) as a free parameter to be constrained by the data. It is convenient to define the dimensionless parameter $Q=q\rho_0$, where $\rho_0$ is the present-day critical density. 

\begin{figure}
\begin{center}
\includegraphics[width=\columnwidth,keepaspectratio]{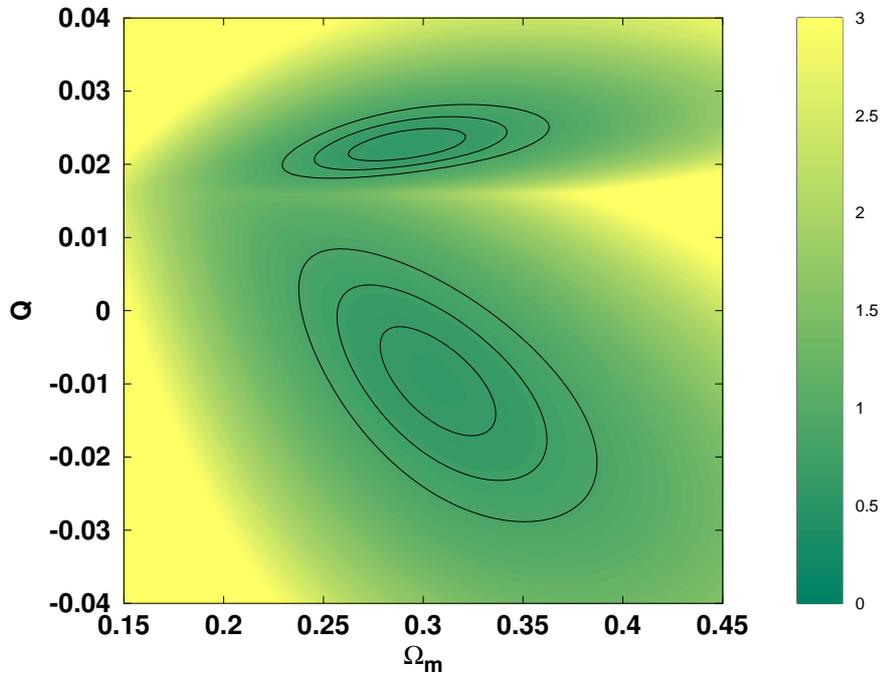}
\end{center}
\caption{Two-dimensional constraints on the $w=0$ generalized coupling model. The $\Delta\chi^2=2.3$, $\Delta\chi^2=6.17$ and $\Delta\chi^2=11.8$ confidence levels are shown in black lines, and the color map depicts the reduced chi-square at each point in the parameter space, with points with $\chi^2_\nu>3$ shown in yellow. Similar constraints can be found in Figure 3 of Ref. \citenum{Fernandes}.}
\label{fig2}
\end{figure}

Since the model effectively has two types of vacuum energy, viz. the one generated by matter fields as well as the usual cosmological constant, one may wonder if the former is sufficient to yield an accelerating universe without invoking the latter. However, it is again simple to show that this can't be the case \cite{Fernandes}, since in that case the minimum density would be $\Omega_m\sim0.86$; clearly such high matter density universes would be incompatible with observations. Thus in what follows we treat this model as a phenomenological extension of $\Lambda$CDM, with the vacuum energy density of matter fields, $Q$, being an additional model parameter which we now constrain.

In the simpler case where the matter equation of state parameter has the standard value, $w=0$, and agnostically allowing both positive and negative values of the model parameter $Q$, we find that while non-zero values of $Q$ are preferred, the standard value is not significantly excluded, as shown in Fig. \ref{fig2}. We note the existence of two branches of the solution, one with $Q>0$ and the other with $Q<0$, with the former branch being slightly preferred. If we restrict the analysis to the range $Q\le0$, the one-sigma constraints on the two model parameters are
\begin{eqnarray}
\Omega_{m-} &=& 0.31\pm 0.02\\
Q_-&=&-0.010\pm0.006\,;
\end{eqnarray}
conversely, if we restrict the analysis to the range $Q\ge0$ we find
\begin{eqnarray}
\Omega_{m+} &=& 0.29\pm 0.02\\
Q_+&=&0.023\pm0.003\,.
\end{eqnarray}
In all cases the reduced chi-square at the best fit is $\chi^2_\nu\sim0.6$, so the model is clearly overfitting the data. All in all, there is no strong evidence for a non-zero $Q$.

In the general case, allowing the dark energy equation of state parameter $w$ to become a further free parameter, one obtains the following one-sigma constraints on the parameters
\begin{eqnarray}
    \Omega_m &=& 0.29^{-0.09}_{+0.07}\\
    Q &=& -0.018^{+0.005}_{-0.004}\\
    w &=& -0.06^{+0.17}_{-0.08}\,.
\end{eqnarray}
The constraints on the matter density are now significantly weaker, but the two $Q$ branches of the solution are still manifest, as are the degeneracies between the model parameters, as can be seen in Ref. \citenum{Fernandes}. In this case the negative branch is also the preferred one. However, we should also point out that the matter equation of state parameter is already more tightly constrained than this (and this comment also applies to the above constraints on the torsion model). Recent analyses \cite{Thomas,Tutusaus} constrain it, conservatively, to $|w|<0.003$. Using this as a Gaussian prior and repeating the analysis, we recover the constraints on $Q$ and $\Omega_m$ reported above for the $w=0$ case, while the posterior for $w$ itself simply recovers the prior.

\section{Scale invariance: the specific Maeder model}

Maeder's proposed scale invariant model \cite{Maeder} is a specific case of the scale-covariant theory of Canuto \textit{et al.} \cite{Canuto1,Canuto2}. It is well known that the effects of scale invariance disappear upon the presence of matter; the assumption underlying scale invariant models is that at large (i.e., cosmological) scales empty space should still be scale invariant. This again leads to a bimetric theory, with a function $\lambda$ (not to be confused with the parameters introduced in previous sections) playing the role of a scale transformation factor relating the ordinary matter frame to another frame which one assumes to still be scale invariant.

In this case, and with the further assumption of a flat homogeneous and isotropic universe, the Friedmann, Raychaudhuri, and continuity equations are \cite{Canuto1,Canuto2}
\begin{eqnarray}
\left(\frac{\dot a}{a}+\frac{\dot\lambda}{\lambda}\right)^2+\frac{k}{a^2}&=&\frac{1}{3}(\kappa\rho+\Lambda\lambda^2)\\
\frac{\ddot a}{a}+\frac{\ddot \lambda}{\lambda}+\frac{\dot \lambda}{\lambda}\frac{\dot a}{a}-\frac{{\dot\lambda}^2}{\lambda^2}&=&-\frac{\kappa}{6}(\rho+3p-2\Lambda\lambda^2)\\
{\dot\rho}+3(\rho+p)\frac{\dot a}{a}&=&-(\rho+3p)\frac{\dot \lambda}{\lambda}\,,
\end{eqnarray}
which match the standard equations if one chooses $\lambda=1$. Note that for a homogeneous and isotropic model $\lambda$ depends only on time, as does the scale factor.

The recent work of Maeder further postulates that the Minkowski metric is a solution of these Einstein equations, which leads to the following consistency conditions \cite{Maeder}
\begin{eqnarray}
3\frac{{\dot\lambda}^2}{\lambda^2}&=&\Lambda\lambda^2\\
2\frac{\ddot \lambda}{\lambda}-\frac{{\dot\lambda}^2}{\lambda^2}&=&\Lambda\lambda^2\,,
\end{eqnarray}
and further imply that
\begin{equation}
\lambda(t)=\sqrt{\frac{3}{\Lambda}}\,\frac{1}{t}\,.
\end{equation}
We are again using $c=1$, and constant equations of state, $p=w\rho$. Together with the solution for $\lambda$, the continuity equation yields
\begin{equation}
\rho\propto (1+z)^{3(1+w)}t^{1+3w}\,;     
\end{equation}
For a cosmological constant equation of state ($w=-1$) this becomes $\rho\propto t^{-2}$; in other words, this is effectively a model with a time-dependent cosmological constant, but no parametric $\Lambda$CDM limit. The author claims  \cite{Maeder}, from a simple qualitative comparison, that with the choice $\Omega_m=0.3$ the model is in good agreement with Hubble parameter data. In Ref. \citenum{Fernandes} this claim was assessed with a more thorough statistical analysis, and we summarize the results here.

\begin{figure}
\begin{center}
\includegraphics[width=\columnwidth,keepaspectratio]{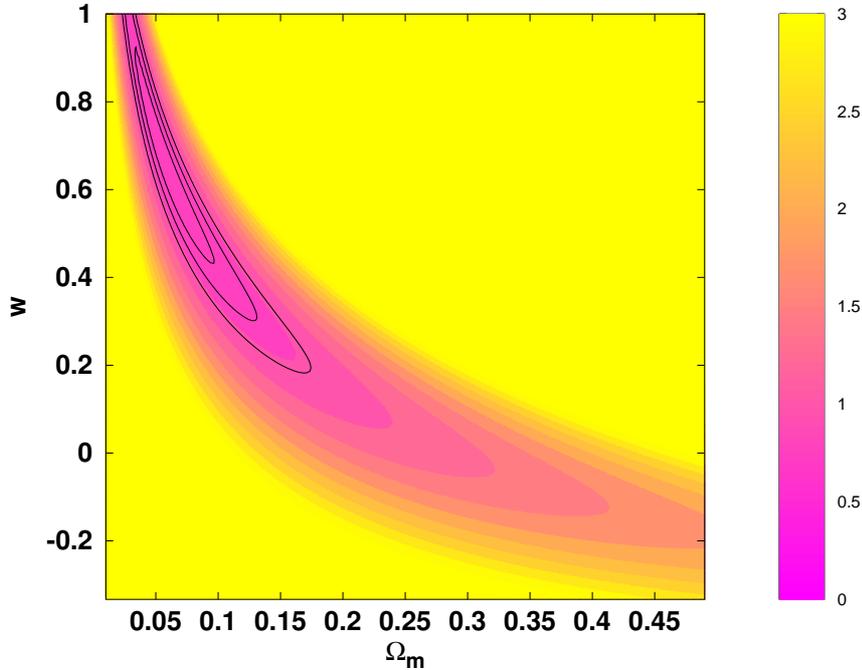}
\end{center}
\caption{Two-dimensional constraints on the Maeder model, with $w$ as a free parameter. The black lines represent the one, two and three sigma confidence levels, and the color map depicts the reduced chi-square at each point in the parameter space, with points with $\chi^2_\nu>3$ shown in yellow. Similar constraints can be found in Figure 6 of Ref. \citenum{Fernandes}.}
\label{fig3}
\end{figure}

With the aforementioned assumptions, the Friedmann equation for the Maeder model can be written
\begin{equation}
E^2(z,x)=\Omega_m(1+z)^{3(1+w)}x^{1+3w}+\frac{\Omega_\lambda}{x}E(z,x)\,,
\end{equation}
where we have defined an effective parameter
\begin{equation}
\Omega_\lambda=\frac{2}{t_0H_0}\,,
\end{equation}
which effectively quantifies the present age of the universe (in dimensionless units) and for convenience also introduced a dimensionless time $x=t/t_0$, with $t_0$ being the current age of the universe. With these definitions the Friedmann equation can be re-written in the simpler form
\begin{eqnarray}
E(z,x)&=&\frac{\Omega_\lambda}{2x}\left[]1+\sqrt{1+M(z,x)}\right]\\
M(z,x)&=&\frac{4\Omega_m}{\Omega_\lambda^2}(1+z)^{3(1+w)}x^{3(1+w)}\,,
\end{eqnarray}
with the relation between redshift and (dimensionless) time being given by
\begin{equation}
\frac{dx}{dz}=- \frac{x}{1+z}\times\frac{1}{1+\sqrt{1+M(z,x)}}\,
\end{equation}
and the initial condition $x=1$ at $z=0$.

In the $w=0$ case we can write $\Omega_\lambda=1-\Omega_m$. and the one-sigma posterior constraint in the matter density is
\begin{equation}
\Omega_m=0.26\pm0.02\,,\quad \chi^2_\nu=1.3\,;
\end{equation}
the inclusion of curvature as an additional parameter \cite{Fernandes} slightly increases the preferred matter density but provides an equally poor fit. On the other hand, allowing $w$ as a further free parameter, one obtains the result shown in in Fig. \ref{fig3}. In this case the one sigma constraints are
\begin{eqnarray}
\Omega_m &=& 0.06\pm0.02\\ 
w &=&0.60^{+0.16}_{-0.15}\,;
\end{eqnarray}
again the inclusion of curvature does not significantly change this \cite{Fernandes}. In both cases the reduced chi-square is now $\chi^2_\nu=0.8$, so the model is now slightly overfitting the data. Clearly there is a strong degeneracy between the matter density and the equation of state parameter (which are anticorrelated), and the best fit values of both parameters are very far from the standard $\Lambda$CDM ones.

\section{Scale invariance: the general model}

The previous section shows that the Maeder model is ruled out. One may therefore ask whether this conclusion extends to the more general model of Canute \textit{et al.}, also introduced in the previous section. Here we present a very preliminary analysis of this issue. We will assume a generic power-law behaviour, $\lambda(t)\propto t^p$, choosing $\lambda_0=1$, and further assuming flat models\footnote{ We will report on a more detailed analysis, relaxing some of these assumptions and exploring various scenarios, elsewhere.}. This choice of $\lambda$ also ensures that $\Lambda$CDM is recovered for $p=0$.

Note that in the Maeder model there is no explicit cosmological constant $\Lambda$. In the general case it is still there, so we may again expect two classes of solutions. One has the usual $\Lambda$ providing the acceleration, with the $\lambda$ field providing a further contribution; in other words, this will be an extension of $\Lambda$CDM. The other has $\Lambda=0$, meaning that the model will not have a $\Lambda$CDM limit, and the question is then whether the field $\lambda$ can provide an alternative to acceleration in that case.

In this case the continuity equation gives
\begin{equation}
\rho\propto (1+z)^{3(1+w)}t^{-p(1+3w)}\,,
\end{equation}
while the Friedmann equation gives
\begin{equation}
\left(E(z,x)+\frac{p}{2x}\Omega_\lambda\right)^2=\Omega_m(1+z)^{3(1+w)}x^{-p(1+3w)}+\Omega_{\Lambda}x^{2p}\,.
\label{scalef}
\end{equation}
For $p=-1$ the Maeder model is recovered with the further assumption that
\begin{equation}
\Omega_\Lambda=\frac{1}{4}\Omega^2_\lambda=\frac{1}{(t_0H_0)^2}\,.
\end{equation}
In the general case the Friedmann equation can be re-written
\begin{equation}
E(z,x)=\frac{\Omega_\lambda}{2x}\left[-p+\sqrt{N(z,x)}\right]
\end{equation}
\begin{equation}
N(z,x)=\frac{4}{\Omega_\lambda^2}\left[\Omega_m(1+z)^{3(1+w)}x^{2-p(1+3w)}+\Omega_\Lambda x^{2(1+p)}\right]\,,
\end{equation}
and the relation between redshift and (dimensionless) time is now given by
\begin{equation}
\frac{dx}{dz}=- \frac{x}{1+z}\times\frac{1}{\sqrt{N(z,x)}-p}\,,
\end{equation}
with the initial condition still being $x=1$ at $z=0$. One can easily check that the Maeder model equations are recovered in the appropriate limit. Note that $\Omega_\lambda$ is a dimensionless measure of the current age of the universe, and it must therefore be a positive quantity.

\begin{figure}
\begin{center}
\includegraphics[width=\columnwidth,keepaspectratio]{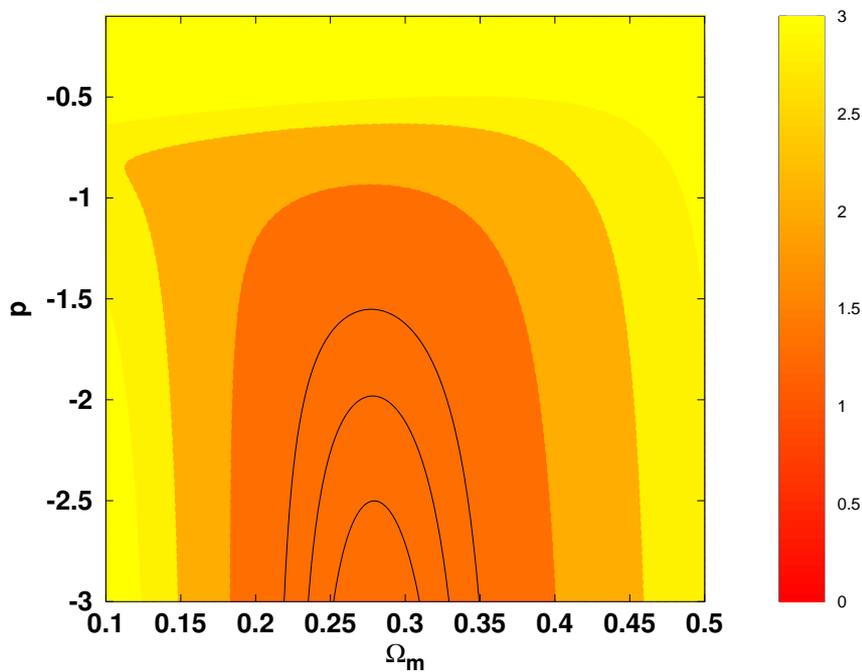}
\end{center}
\caption{Two-dimensional constraints on the Canuto \textit{et al.} model, with $w=0$ and $\Omega_\Lambda=0$. The black lines represent the one, two and three sigma confidence levels, and the color map depicts the reduced chi-square at each point in the parameter space, with points with $\chi^2_\nu>3$ shown in yellow.}
\label{fig4}
\end{figure}

Several different scenarios can now be considered. In what follows we only consider the simplest one. If we assume that $\Omega_\Lambda=0$ and ordinary matter with an $w=0$ equation of state we have two free parameters ($\Omega_m,p$) and
\begin{equation}
-p\Omega_\lambda=2(1-\sqrt{\Omega_m})\,;    
\end{equation}
thus $p=0$ corresponds to an $\Omega_m=1$ universe, In this case we again find that this model does not provide a good fit to the data, cf. Fig. \ref{fig4}: while a matter density of around $\Omega_m\sim0.28$ is preferred, the reduced chi-square of the best fit is quite poor, being always larger than $\chi^2_\nu=1.25$. An exploration of the wider parameter space, allowing for a cosmological constant, a non-zero equation of state, and curvature, will be reported elsewhere.

\section{Outlook}

We have briefly presented a comparison of three classes of models for the low-redshift acceleration of the universe against background low-redshift cosmological observations, further using the traditional CPL phenomenological parameterization as a benchmark. Each of these models contains, in principle, an additional mechanism, in addition to the cosmological constant, that could account for the recent acceleration of the universe.

We find that the steady-state torsion and generalized couplings have similar behaviours. The specific physical mechanism therein are ruled out as unique source of acceleration, but small (e.g. percent level) deviations from $\Lambda$CDM are still allowed by the data that we have considered. Constraints can of course be tightened by including further data. The scale invariant model provides an interesting contrast. Assuming the standard equation of state parameter, $w=0$, the best-fit value is similar to the CPL one. However, these fits have a poor (specifically, high) reduced chi-square, indicating that the model does not provide a good fit to the data. An observationally viable model of this kind  would require a highly non-standard ‘matter’ density (including a non-standard equation of state), which would conflict with other cosmological datasets. In passing, we also mention that another interesting phenomenological class is that of energy-momentum-powered models recently constrained in Ref. \citenum{Faria} and also discussed elsewhere in these proceedings.

Overall, we therefore conclude that $\Lambda$CDM is a remarkably robust paradigm. While it is clearly a phenomenological approximation to a still unknown more fundamental model, it is clearly a good one, and any plausible alternative model must be able to closely reproduce its behaviour in a broad range of cosmological settings.

\section*{Acknowledgments}

This work was financed by FEDER---Fundo Europeu de Desenvolvimento Regional funds through the COMPETE 2020---Operational Programme for Competitiveness and Internationalisation (POCI), and by Portuguese funds through FCT - Funda\c c\~ao para a Ci\^encia e a Tecnologia in the framework of the project POCI-01-0145-FEDER-028987 and PTDC/FIS-AST/28987/2017.

\eject

\bibliographystyle{ws-procs961x669}
\bibliography{martinsalternatives}

\end{document}